\documentclass[proceedings]{JHEP} 

\usepackage{epsfig}			

\newbox\mybox
\newcommand\fverb{\setbox\mybox=\hbox\bgroup\verb}
\newcommand\fverbdo{\egroup\medskip\noindent\fbox{\unhbox\mybox}\ }
\newcommand\fverbit{\egroup\item[\fbox{\unhbox\mybox}]}

\font\beeg=cmr17 scaled 1600		
\newcommand\init[1]{\setbox\mybox=\hbox{{\beeg #1}~}%
		   \noindent\global\hangindent=\wd\mybox\global\hangafter-2%
		   \sc\smash{\llap {\lower 13.2pt \box\mybox}}}

\title{Exclusive semileptonic \boldmath{B} decays to excited \boldmath{D} mesons}

\author{D.\ Ebert$^\dag$, R.\ N.\ Faustov$^\ddag$, V.\ O.\ Galkin$^\ddag$\\
$^\dag$ Institut f\"ur Physik, Humboldt--Universit\"at zu Berlin,
Invalidenstr.110, D-10115 Berlin, Germany\\
$^\ddag$ Russian Academy of Sciences,
Scientific Council for Cybernetics,
Vavilov Street 40, Moscow 117333, Russia}

\conference{Heavy Quark Physics 5, Dubna, Russia, 6-8 April 2000}

\abstract{Exclusive semileptonic $B$ decays to orbitally and 
radially excited charmed mesons
are investigated in the first order of the heavy quark expansion.
The emerging leading and subleading Isgur-Wise functions are calculated
in the framework of the relativistic quark model. It is found that
both relativistic and the $1/m_Q$ corrections play an important role and substantially
modify results. An interesting interplay between different corrections
is observed. }

\begin{document} 
\section{Introduction}
The investigation of  semileptonic decays of $B$ mesons to excited
charmed mesons represents a problem interesting both from the experimental and
theoretical point of view. The current experimental data on semileptonic
$B$ decays to ground state $D$ mesons indicate that a substantial part  
($\approx 40\%$) of the inclusive semileptonic $B$ decays should go to 
excited $D$ meson states. First experimental data on some 
exclusive $B$ decay channels to excited charmed mesons are becoming 
available now \cite{cleo,aleph,opal} and more data are expected in near
future. Thus the comprehensive theoretical study of these decays is 
necessary. The presence of the heavy 
quark in the initial and final meson states in these decays considerably
simplifies their theoretical description. A good starting point for this
analysis is the infinitely heavy quark limit, $m_Q\to\infty$ \cite{iw}. 
In this limit the heavy quark symmetry arises, which strongly
reduces the number of independent weak form factors \cite{iw1}.
The heavy quark mass and spin then decouple and all meson properties are
determined by light-quark degrees of freedom alone. This leads to
a considerable reduction of the number of independent form factors
which are necessary for the description of heavy-to-heavy semileptonic
decays. For example, in this limit only one form factor is necessary 
for the semileptonic
$B$ decay to $S$-wave $D$ mesons (both for the ground state and its
radial excitations), while the decays to $P$ states require two form
factors \cite{iw1}. It is important to note 
that in the infinitely heavy quark limit matrix elements between
a $B$ meson and an excited $D$ meson should vanish at the point of
zero recoil of the final excited charmed meson in the rest frame of the $B$
meson. In the case of $B$ decays to radially excited charmed mesons
this follows from the orthogonality of radial parts of wave functions,
while for the decays to orbital excitations this is the consequence of
orthogonality of their angular parts. However, some of the $1/m_Q$ corrections
to these decay matrix elements can give nonzero contributions at zero 
recoil. As a result the role of these  corrections could be considerably
enhanced, since the kinematical range for $B$ decays to excited states is
a rather small region around zero recoil. 

Our relativistic quark model is  based on the quasipotential approach in
quantum field
theory with a specific choice  of the quark-antiquark interaction
potential. It provides
a consistent scheme for the  calculation of all relativistic corrections
at a given $v^2/c^2$ 
order and allows for the heavy  quark $1/m_Q$ expansion. In preceding
papers 
we applied this model to the calculation of the mass spectra of 
orbitally and radially excited states of heavy-light mesons \cite{egf},
as well as to a description of weak decays of $B$ mesons to ground state
heavy and light mesons \cite{fgm,efg}. The heavy quark expansion for the ground
state heavy-to-heavy semileptonic transitions \cite{fg} was found to be 
in agreement with model-independent predictions of the heavy quark effective theory 
(HQET).

  \section{Decay matrix elements and the heavy quark
      expansion}
In this section we present the heavy quark expansion  for weak decay 
matrix elements between a $B$ meson and radially excited charmed meson
states up to the first order in $1/m_Q$  using  the HQET. The  corresponding formulars
for $B$ decays to orbital excitations can be found in Ref.~\cite{llsw}.
       
The matrix elements of the vector  and axial vector currents  between
$B$ and radially excited $D'$ or $D^{*}{'}$ mesons can be parameterized by six
hadronic form factors:
 \begin{eqnarray}\label{ff}
 &&\!\!\!\!\!\!\!{\langle D'(v')| \bar c\gamma^\mu b |B(v)\rangle  
\over\sqrt{m_{D'}m_B}}
  = h_+ (v+v')^\mu + h_- (v-v')^\mu , \cr
 &&\!\!\!\!\!\!\! \langle D'(v')| \bar c\gamma^\mu  \gamma_5 b|B(v)\rangle 
  = 0, \cr
&&\!\!\!\!\!\!\!  {\langle D^*{'}(v',\epsilon)| \bar c\gamma^\mu b |B(v)\rangle  
\over\sqrt{m_{D^*{'}}m_B}}
  = i h_V \varepsilon^{\mu\alpha\beta\gamma} 
  \epsilon^*_\alpha v'_\beta v_\gamma ,\cr
&&\!\!\!\!\!\!\!  {\langle D^*{'}(v',\epsilon)| \bar c\gamma^\mu\gamma_5 b |B(v)\rangle  
\over\sqrt{m_{D^*{'}}m_B}}
 = h_{A_1}(w+1) \epsilon^{* \mu} \cr
 &&  -(h_{A_2} v^\mu + h_{A_3} v'^\mu) (\epsilon^*\cdot v) ,
   \end{eqnarray}
where $v~(v')$ is the four-velocity of the $B~(D^{(*)}{'})$ meson,
$\epsilon^\mu$ is a  polarization vector  of the final vector
charmed meson, and the form factors $h_i$  are dimensionless 
functions of the product of velocities $w=v\cdot v'$.    

Now we expand the form factors $h_i$ in powers of $1/m_Q$ up to first
order and relate the coefficients in this expansion to universal Isgur-Wise
functions. This is achieved by evaluating the matrix elements of the
effective current operators arising from the HQET expansion of the weak
currents. For simplicity we limit our analysis to the leading order in
$\alpha_s$ and use the trace formalism \cite{falk}. 
Following Ref.~\cite{llsw}, we introduce the matrix
\begin{equation}
H_v = \frac{1+\not\! v}{2} \Big[ P_v^{*\mu} \gamma_\mu 
  - P_v \gamma_5 \Big],  \label{Hdef}
\end{equation} 
composed from the fields $P_v$ and $P_v^{*\mu}$ that destroy mesons in
the $j^{P}=\frac12^-$ doublet \footnote{Here $j$ is the total light quark 
angular momentum, and the superscript $P$ denotes the meson parity.} 
with four-velocity $v$. At leading
order of the heavy quark expansion ($m_Q\to\infty$) the matrix elements of
the weak current between the  ground and radially excited states 
destroyed  by the fields in $H_v$ and $H'_v$, respectively, 
are given by
\begin{equation}\label{lo}
\bar c \Gamma b \to \bar h^{(c)}_{v'}\Gamma h^{(b)}_v = \xi^{(n)}(w)
  {\rm Tr} \Big\{\bar H'_{v'} \Gamma H_v \Big\},
\end{equation} 
where $h_v^{(Q)}$ is the heavy quark field in the effective theory.
The leading order Isgur-Wise function $\xi^{(n)}(w)$ vanishes at the 
zero recoil ($w=1$) of the final meson for any $\Gamma$, because of the
heavy quark symmetry and the orthogonality of the radially excited state
wave function with respect to the ground state one.
 
At first order of the $1/m_Q$ expansion there are\hfill contributions\hfill from\hfill
the\hfill corrections\hfill to\hfill the\newline HQET Lagrangian 
 \begin{eqnarray}\label{lcor}
&&\!\!\!\!\!\!\!\!\!\!\!\!\!\delta{\cal L} =  \frac1{2m_Q}{\cal L}_{1,v}^{(Q)}
\equiv\frac1{2 m_Q} \Big[ O_{{\rm kin},v}^{(Q)} +
  O_{{\rm mag},v}^{(Q)} \Big], \\
&&O_{{\rm kin},v}^{(Q)} = \bar h_v^{(Q)} (iD)^2 h_v^{(Q)}, \cr
&&O_{{\rm mag},v}^{(Q)} = \bar h_v^{(Q)}
  \frac{g_s}2 \sigma_{\alpha\beta} G^{\alpha\beta} h_v^{(Q)}\nonumber
\end{eqnarray}  
and from the tree-level matching of the weak current operator onto
effective theory which contains a
covariant derivative $D^{\lambda}=\partial^{\lambda}-ig_st_aA_a^{\lambda}$
 \begin{equation}\label{ccor}
\bar c \Gamma b \to \bar h_{v'}^{(c)} 
  \bigg( \Gamma - \frac i{2m_c} \overleftarrow {\not\!\! D} \Gamma  
  + \frac i{2m_b} \Gamma \overrightarrow {\not\!\! D} 
  \bigg) h_v^{(b)}. 
\end{equation}
 The matrix elements of the latter operators can be parameterized as
\begin{eqnarray}\label{cur}
\bar h^{(c)}_{v'} i\overleftarrow D_{\lambda} \Gamma h^{(b)}_v &=&
  {\rm Tr}\Big\{ \xi_{\lambda}^{(c)} 
  \bar H_{v'}\Gamma H_v \Big\} , \cr
\bar h^{(c)}_{v'} \Gamma i\overrightarrow D_{\lambda} h^{(b)}_v &=&
  {\rm Tr} \Big\{  \xi^{(b)}_{\lambda} 
  \bar H_{v'} \Gamma H_v \Big\}.  
\end{eqnarray}
The most general form for $\xi^{(Q)}_{\lambda}$ is \cite{luke}
\begin{equation}
\xi^{(Q)}_{\lambda}=\xi^{(Q)}_{+}(v+v')_{\lambda}+\xi^{(Q)}_{-}
(v-v')_{\lambda}-\xi^{(Q)}_3\gamma_{\lambda}.
\end{equation}
The \hfill equation\hfill of\hfill motion\hfill for\hfill the\hfill heavy\hfill 
quark,\newline\mbox{$i(v\cdot D)h^{(Q)}=0$}, yields
the relations between the form factors $\xi^{(Q)}_i$
\begin{eqnarray}\label{rel1}
\xi_{+}^{(c)}(1+w)+\xi_{-}^{(c)}(w-1)+\xi_3^{(c)}&=&0\cr
\xi_{+}^{(b)}(1+w)-\xi_{-}^{(b)}(w-1)+\xi_3^{(b)}&=&0.
\end{eqnarray}
The additional relations can be obtained from the  momentum conservation
and the definition of the heavy quark fields $h_v^{(Q)}$, which lead to
the equation 
$ i\partial_\nu(\bar h_{v'}^{(c)}\Gamma\,h_v^{(b)}) = 
  (\bar\Lambda v_\nu-\bar\Lambda^{(n)}v'_\nu)\bar h_{v'}^{(c)}\Gamma 
h_v^{(b)}$, implying that
\begin{equation}\label{cnst}
{\xi}^{(c)}_{\lambda} + {\xi}^{(b)}_{\lambda}
  = (\bar\Lambda v_\lambda-\bar\Lambda^{(n)}v'_\lambda) \xi^{(n)}.
\end{equation}
Here $\bar\Lambda(\bar\Lambda^{(n)})=M(M^{(n)})-m_Q$ is the difference between
the heavy ground state (radially excited) meson and heavy quark masses in 
the limit $m_Q\to\infty$. This equation results in the following relations
\begin{eqnarray}\label{rel2}
\xi^{(c)}_{+}+\xi^{(b)}_{+}+\xi^{(c)}_{-}+\xi^{(b)}_{-}
&=&\bar\Lambda\xi^{(n)},\cr
\xi^{(c)}_{+}+\xi^{(b)}_{+}-\xi^{(c)}_{-}-\xi^{(b)}_{-}
&=&-\bar\Lambda^{(n)}\xi^{(n)},\cr
\xi_3^{(c)}+\xi_3^{(b)}&=&0.
\end{eqnarray}
The relations (\ref{rel1}) and (\ref{rel2}) can be used to 
express the functions
$\xi_{-,+}^{(Q)}$ in terms of $\tilde\xi_3(\equiv\xi_3^{(b)}=-\xi_3^{(c)})$ and
the leading order function $\xi^{(n)}$:
\begin{eqnarray}\label{corc}
\!\!\!\!\!\!\xi^{(c)}_{-}&=&\left(\frac{\bar\Lambda^{(n)}}2+
\frac12\frac{\bar\Lambda^{(n)}-\bar\Lambda}{w-1}\right)\xi^{(n)},\cr
\!\!\!\!\!\!\xi^{(b)}_{-}&=&\left(\frac{\bar\Lambda}2-
\frac12\frac{\bar\Lambda^{(n)}-\bar\Lambda}{w-1}\right)\xi^{(n)},\cr
\!\!\!\!\!\!\xi^{(c)}_{+}&=&\left(-\frac{\bar\Lambda^{(n)}}2+
\frac12\frac{\bar\Lambda^{(n)}+\bar\Lambda}{w+1}\right)\xi^{(n)}+
\frac1{w+1}\tilde\xi_3,\cr
\!\!\!\!\!\!\xi^{(b)}_{+}&=&\left(\frac{\bar\Lambda}2-
\frac12\frac{\bar\Lambda^{(n)}+\bar\Lambda}{w+1}\right)\xi^{(n)}-
\frac1{w+1}\tilde\xi_3.
\end{eqnarray}

The matrix elements of the $1/m_Q$ corrections resulting from insertions
of higher-dimension operators of the HQET Lagrangian (\ref{lcor}) 
have the structure \cite{luke}
\begin{eqnarray}\label{corl}
&&\!\!\!\!\!\!\!\!\!\!\!\!\!\!i \int {\rm d}x\, T\left\{ {\cal L}_{1,v'}^{(c)}(x) 
  \left[ \bar h_{v'}^{(c)} \Gamma h_{v}^{(b)} \right](0) \right\} \cr
&&  = 2\chi^{(c)}_1 {\rm Tr} \left\{\bar H_{v'} \Gamma H_v \right\}\cr
&&+
2{\rm Tr}\left\{ \chi_{\alpha\beta}^{(c)}
  \bar H_{v'} i\sigma^{\alpha\beta} \frac{1+\not \! v'}2 
  \Gamma H_v \right\},\cr
&&\!\!\!\!\!\!\!\!\!\!\!\!\!\!i \int {\rm d}x\, T\left\{ {\cal L}_{1,v}^{(b)}(x) 
  \left[ \bar h_{v'}^{(c)} \Gamma h_{v}^{(b)} \right](0) \right\} \cr
 & &= 2\chi^{(b)}_1 {\rm Tr} \left\{\bar H_{v'} \Gamma H_v \right\}\cr
&&+
2{\rm Tr}\left\{ \chi_{\alpha\beta}^{(b)}
  \bar H_{v'}\Gamma \frac{1+\not\! v}2 i\sigma^{\alpha\beta}  
   H_v \right\}.
\end{eqnarray}
The corrections coming from the kinetic energy term $O_{\rm kin}$ 
do not violate spin symmetry and, hence, the corresponding functions
$\chi_1^{(Q)}$ effectively correct the leading order function $\xi^{(n)}$.
The chromomagnetic\hfill operator\hfill $O_{\rm mag}$,\hfill on\hfill the\hfill other
\newline hand, explicitly
violates spin symmetry. The most general decomposition of the tensor
form factor $\chi_{\alpha\beta}^{(Q)}$ is \cite{luke,n}
\begin{eqnarray}
\chi_{\alpha\beta}^{(c)}&=&\chi_2^{(c)}v_{\alpha}\gamma_{\beta} 
-\chi_3^{(c)}i\sigma_{\alpha\beta},\cr
\chi_{\alpha\beta}^{(b)}&=&\chi_2^{(b)}v'_{\alpha}\gamma_{\beta} 
-\chi_3^{(b)}i\sigma_{\alpha\beta}.
\end{eqnarray}

The functions $\chi_i^{(b)}$ contribute to the decay form factors (\ref{ff})
only in the linear combination $\chi_b=2\chi_1^{(b)}-4(w-1)\chi_2^{(b)}
+12\chi_3^{(b)}$. Thus five independent functions $\tilde\xi_3$, $\chi_b$ and
$\tilde\chi_i(\equiv \chi_i^{(c)})$, as well as two mass parameters
$\bar\Lambda$ and $\bar\Lambda^{(n)}$ are necessary 
to describe first order $1/m_Q$
corrections to  matrix elements of $B$ meson decays to radially excited
$D$ meson states. The resulting structure of the decay form factors is
\begin{eqnarray}\label{cff}
&&\!\!\!\!\!\!\!h_{+}=\xi^{(n)}+\varepsilon_c\left[2\tilde\chi_1-4(w-1)\tilde\chi_2+
12\tilde\chi_3\right]+\varepsilon_b\chi_b,\cr
&&\!\!\!\!\!\!\!h_{-}=\varepsilon_c\left[2\tilde\xi_3-\left(\bar\Lambda^{(n)}+
\frac{\bar\Lambda^{(n)}-\bar\Lambda}{w-1}\right)\xi^{(n)}\right]\cr
&&\!\!\!\!\!\!\!- \varepsilon_b\left[2\tilde\xi_3-\left(\bar\Lambda-
\frac{\bar\Lambda^{(n)}-\bar\Lambda}{w-1}\right)\xi^{(n)}\right],\cr
&&\!\!\!\!\!\!\!h_V=\xi^{(n)}+\varepsilon_c\biggl[2\tilde\chi_1+
\left(\bar\Lambda^{(n)}+\frac{\bar\Lambda^{(n)}
-\bar\Lambda}{w-1}\right)\xi^{(n)}\cr
&&\!\!\!\!\!\!\!-4\tilde\chi_3\biggr]+\varepsilon_b\left[\chi_b+
\left(\bar\Lambda-\frac{\bar\Lambda^{(n)}-\bar\Lambda}{w-1}\right)\xi^{(n)}
-2\tilde\xi_3\right],\cr
&&\!\!\!\!\!\!\!h_{A_1}=\xi^{(n)}+\varepsilon_c\biggl[2\tilde\chi_1-4\tilde\chi_3\cr
&&\!\!\!\!\!\!\!+
\frac{w-1}{w+1}\left(\bar\Lambda^{(n)}+\frac{\bar\Lambda^{(n)}
-\bar\Lambda}{w-1}\biggr)\xi^{(n)}\right]\cr
&&\!\!\!\!\!\!\!+\varepsilon_b\left\{\chi_b+
\frac{w-1}{w+1}\left[
\left(\bar\Lambda-\frac{\bar\Lambda^{(n)}-\bar\Lambda}{w-1}\right)\xi^{(n)}
-2\tilde\xi_3\right]\right\},\cr
&&\!\!\!\!\!\!\!h_{A_2}=\varepsilon_c\biggl\{4\tilde\chi_2-\frac2{w+1}\biggl[
\left(\bar\Lambda^{(n)}+\frac{\bar\Lambda^{(n)}
-\bar\Lambda}{w-1}\right)\xi^{(n)}\cr
&&+\tilde\xi_3\biggr]\biggr\},\cr
&&\!\!\!\!\!\!\!h_{A_3}=\xi^{(n)}+\varepsilon_c\biggl[2\tilde\chi_1-4\tilde\chi_2-
4\tilde\chi_3\cr
&&\!\!\!\!\!\!\!+\frac{w-1}{w+1}\left(\bar\Lambda^{(n)}+\frac{\bar\Lambda^{(n)}
-\bar\Lambda}{w-1}\right)\xi^{(n)}-\frac2{w+1}\tilde\xi_3\biggr]\cr
&&\!\!\!\!\!\!\!+\varepsilon_b\left[\chi_b+
\left(\bar\Lambda-\frac{\bar\Lambda^{(n)}-\bar\Lambda}{w-1}\right)\xi^{(n)}
-2\tilde\xi_3\right],
\end{eqnarray}
where $\varepsilon_Q=1/(2m_Q)$.

The similar analysis \cite{llsw} for $B$ decays to orbitally excited states indicate
that it is necessary to introduce two Isgur-Wise functions in leading order of the
heavy quark expansion: one function $\tau(w)$ for decays to $D_1,D^*_2$ mesons 
with $j=3/2$  and the other one $\zeta(w)$ for decays to $D^*_0,D^*_1$ mesons with 
$j=1/2$. At subleading order six additional functions ($\tau_{1,2}$, $\eta_{ke}$, 
$\eta_{1,2,3}$) arise for the former decays and four functions ($\zeta_1$, $\chi_{ke}$,
$\chi_{1,2}$) for the latter ones.

\section{Relativistic quark model}

We use the relativistic quark model based on the quasipotentil approach for the
calculation of corresponding Isgur-Wise functions. Our model has been described
in detail at this conference \cite{mconf}, so we directly go to the calculation of 
decay matrix elements of the weak current between meson states.
In the quasipotential approach,  the matrix element of the weak current
$J^W=\bar c\gamma_\mu(1-\gamma^5)b$ 
between a $B$ meson and an  excited $D^{**}$ meson takes
 the form \cite{f}
\begin{eqnarray}\label{mxet} 
&&\!\!\!\!\!\!\!\!\!\!\!\!\!\!\langle D^{**} \vert J^W_\mu (0) \vert B\rangle\cr
&&\!\!\!=\int \frac{d^3p\, d^3q}{(2\pi )^6} \bar \Psi_{D^{**}}({\bf
p})\Gamma _\mu ({\bf p},{\bf q})\Psi_B({\bf q}),\end{eqnarray}
where $\Gamma _\mu ({\bf p},{\bf
q})$ is the two-particle vertex function and  $\Psi_{B,D^{**}}$ are the
meson wave functions projected onto the positive energy states of
quarks and boosted to the moving reference frame.
 The contributions to $\Gamma$ come from Figs.~1 and 2.\footnote{  
The contribution $\Gamma^{(2)}$ is the consequence
of the projection onto the positive-energy states. Note that the form of the
relativistic corrections resulting from the vertex function
$\Gamma^{(2)}$ is explicitly dependent on the Lorentz structure of the
$q\bar q$-interaction.} In the heavy quark limit
$m_{b,c}\to \infty$ only $\Gamma^{(1)}$ contributes, while $\Gamma^{(2)}$ 
contributes at $1/m_{Q}$ order. 
They look like
\begin{equation} \label{gamma1}
\!\Gamma_\mu^{(1)}({\bf
p},{\bf q})=\bar u_{c}(p_c)\gamma_\mu(1-\gamma^5)u_b(q_b)
(2\pi)^3\delta({\bf p}_q-{\bf
q}_q),\end{equation}
and
\begin{eqnarray}\label{gamma2} 
&&\!\!\!\!\!\!\!\Gamma_\mu^{(2)}({\bf
p},{\bf q})=\bar u_{c}(p_c)\bar u_q(p_q) \Bigl\{\gamma_{Q\mu}(1-\gamma_Q^5)\cr
&&\times
\frac{\Lambda_b^{(-)}(
k)}{\epsilon_b(k)+\epsilon_b(p_c)}\gamma_Q^0
{\cal V}({\bf p}_q-{\bf
q}_q) 
+{\cal V}({\bf p}_q-{\bf
q}_q)\cr
&&\times\frac{\Lambda_{c}^{(-)}(k')}{ \epsilon_{c}(k')+
\epsilon_{c}(q_b)}
\gamma_Q^0 \gamma_{Q\mu}(1-\gamma_Q^5)\Bigr\}u_b(q_b)
u_q(q_q),\cr&&
\end{eqnarray}
where the superscripts ``(1)" and ``(2)" correspond to Figs.~1 and
2, $Q= c$ or $b$, ${\bf k}={\bf p}_c-{\bf\Delta};\
{\bf k}'={\bf q}_b+{\bf\Delta};\ {\bf\Delta}={\bf
p}_{D^{**}}-{\bf p}_B; \ \epsilon (p)=(m^2+{\bf p}^2)^{1/2}$;
$$\Lambda^{(-)}(p)=\frac{\epsilon(p)-\bigl( m\gamma
^0+\gamma^0(\mbox{\boldmath{$\gamma$\bf p}})\bigr)}{ 2\epsilon (p)}.$$

\FIGURE{
\unitlength=0.45mm
\small
\begin{picture}(150,150)
\put(10,100){\line(1,0){50}}
\put(10,120){\line(1,0){50}}
\put(35,120){\circle*{8}}
\multiput(32.5,130)(0,-10){2}{\begin{picture}(5,10)
\put(2.5,10){\oval(5,5)[r]}
\put(2.5,5){\oval(5,5)[l]}\end{picture}}
\put(5,120){$b$}
\put(5,100){$\bar q$}
\put(5,110){$B$}
\put(65,120){$c$}
\put(65,100){$\bar q$}
\put(65,110){$D^{**}$}
\put(43,140){$W$}
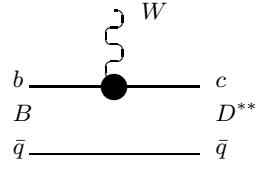
\put(-7,85){\makebox[7.3cm][s]{\small {\bf Figure 1}: Lowest order vertex function $\Gamma^{(1)}$
cont-}}
\put(-7,76){ributing to the current matrix element (\ref{mxet}). }
\put(10,20){\line(1,0){50}}
\put(10,40){\line(1,0){50}}
\put(25,40){\circle*{8}}
\put(25,40){\thicklines \line(1,0){20}}
\multiput(25,40.5)(0,-0.1){10}{\thicklines \line(1,0){20}}
\put(25,39.5){\thicklines \line(1,0){20}}
\put(45,40){\circle*{2}}
\put(45,20){\circle*{2}}
\multiput(45,40)(0,-4){5}{\line(0,-1){2}}
\multiput(22.5,50)(0,-10){2}{\begin{picture}(5,10)
\put(2.5,10){\oval(5,5)[r]}
\put(2.5,5){\oval(5,5)[l]}\end{picture}}
\put(5,40){$b$}
\put(5,20){$\bar q$}
\put(5,30){$B$}
\put(65,40){$c$}
\put(65,20){$\bar q$}
\put(65,30){$D^{**}$}
\put(33,60){$W$}
\put(90,20){\line(1,0){50}}
\put(90,40){\line(1,0){50}}
\put(125,40){\circle*{8}}
\put(105,40){\thicklines \line(1,0){20}}
\multiput(105,40.5)(0,-0.1){10}{\thicklines \line(1,0){20}}
\put(105,39,5){\thicklines \line(1,0){20}}
\put(105,40){\circle*{2}}
\put(105,20){\circle*{2}}
\multiput(105,40)(0,-4){5}{\line(0,-1){2}}
\multiput(122.5,50)(0,-10){2}{\begin{picture}(5,10)
\put(2.5,10){\oval(5,5)[r]}
\put(2.5,5){\oval(5,5)[l]}\end{picture}}
\put(85,40){$b$}
\put(85,20){$\bar q$}
\put(85,30){$B$}
\put(145,40){$c$}
\put(145,20){$\bar q$}
\put(145,30){$D^{**}$}
\put(133,60){$W$}
\put(-7,5){\makebox[7.3cm][s]{\small {\bf Figure 2}: Vertex function $\Gamma^{(2)}$
taking the quark}} 
\put(-7,-4){\makebox[7.3cm][s]{interaction into account. Dashed lines correspond} }
\put(-7,-13) {\makebox[7.3cm][s]{to the effective potential. Bold lines denote the ne-}}
\put(-7,-22){{gative-energy part of the quark
propagator. }}
\end{picture}
\vspace*{0.7cm}
}
\setcounter{figure}2

\noindent Here \cite{f} 
\begin{eqnarray*} 
p_{c,q}&=&\epsilon_{c,q}(p)\frac{p_{D^{**}}}{M_{D^{**}}}
\pm\sum_{i=1}^3 n^{(i)}(p_{D^{**}})p^i,\\
q_{b,q}&=&\epsilon_{b,q}(q)\frac{p_B}{M_B} \pm \sum_{i=1}^3 n^{(i)}
(p_B)q^i,\end{eqnarray*}
and $n^{(i)}$ are three four-vectors given by
$$ n^{(i)\mu}(p)=\left\{ \frac{p^i}{M},\ \delta_{ij}+
\frac{p^ip^j}{M(E+M)}\right\}.$$

It is important to note that the wave functions entering the weak current
matrix element (\ref{mxet}) are not in the rest frame in general. For example, 
in the $B$ meson rest frame, the $D^{**}$ meson is moving with the recoil
momentum ${\bf \Delta}$. The wave function
of the moving $D^{**}$ meson $\Psi_{D^{**}\,{\bf\Delta}}$ is connected 
with the $D^{**}$ wave function in the rest frame $\Psi_{D^{**}\,{\bf 0}}$
by the transformation \cite{f}
\begin{equation}
\label{wig}
\Psi_{D^{**}\,{\bf\Delta}}({\bf
p})=D_c^{1/2}(R_{L_{\bf\Delta}}^W)D_q^{1/2}(R_{L_{
\bf\Delta}}^W)\Psi_{D^{**}\,{\bf 0}}({\bf p}),
\end{equation}
where $R^W$ is the Wigner rotation, $L_{\bf\Delta}$ is the Lorentz boost
from the meson rest frame to a moving one, and   
the rotation matrix $D^{1/2}(R)$ in spinor representation is given by
\begin{equation}\label{d12}
{1 \ \ \,0\choose 0 \ \ \,1}D^{1/2}_{c,q}(R^W_{L_{\bf\Delta}})=
S^{-1}({\bf p}_{c,q})S({\bf\Delta})S({\bf p}),
\end{equation}
where
$$
S({\bf p})=\sqrt{\frac{\epsilon(p)+m}{2m}}\left(1+\frac{\mbox{\boldmath $\alpha \bf p$}}
{\epsilon(p)+m}\right)
$$
is the usual Lorentz transformation matrix of the four-spinor.
For electroweak $B$ meson
decays to $S$-wave final mesons such a transformation contributes at first
order of the $1/m_Q$ expansion, while for the decays to excited final mesons
it gives a contribution already to the leading term due to the orthogonality
of the initial and final meson wave functions.
 
Now we can perform the heavy quark expansion for the matrix elements
of $B$ decays to excited $D$ mesons in the framework of our model and
determine leading and subleading Isgur--Wise functions. We substitute
the vertex functions $\Gamma^{(1)}$  and $\Gamma^{(2)}$ 
given by Eqs.~(\ref{gamma1}) and (\ref{gamma2})
in the decay matrix element (\ref{mxet})  and take into account the wave
function properties (\ref{wig}).
The resulting structure of this matrix element is
rather complicated, because it is necessary  to integrate both over  $d^3
p$
and $d^3 q$. The $\delta$ function in expression  (\ref{gamma1}) permits
us to perform
one of these integrations and thus this contribution  can be easily
calculated. The
calculation  
of the vertex function $\Gamma^{(2)}$ contribution is  more difficult.
Here, instead
of a $\delta$ function, we have a complicated structure, containing the 
$Q\bar q$ interaction potential in the meson. 
However, we can expand this contribution in inverse
powers of heavy ($b,c$) quark masses and then use the  quasipotential
equation in
order to perform one of the integrations in the current matrix element.  
We carry out 
the heavy quark expansion up to first order in $1/m_Q$. It is easy to see  
that the vertex
function $\Gamma^{(2)}$ contributes already at  the subleading order of  
the $1/m_Q$
expansion. Then we compare the arising  decay matrix elements with
the form factor decompositions (\ref{cff}) for decays to radial excitations 
and the corresponding ones for decays to orbital excitations \cite{llsw} and determine  
the form factors. We find that, for the chosen values of our model parameters  
(the mixing
coefficient of vector and scalar confining potential $\varepsilon=-1$  and
the Pauli
constant $\kappa=-1$), the resulting structure  at leading
and subleading order in $1/m_Q$ coincides with the model-independent  
predictions of HQET. This allows us to determine leading and subleading 
Isgur-Wise functions.
\section{Semileptonic decays to orbitally excited states}  
We get the following
expressions for leading and subleading Isgur--Wise functions of semileptonic
$B$ decays to orbitally excited $D$ mesons \cite{orb}:

(i) $B\to D_1e\nu$ and $B\to D^*_2e\nu$ decays
\begin{eqnarray}
\label{tau}
&&\!\!\!\!\!\!\!\tau(w)=\sqrt{\frac23}\frac{1}{(w+1)^{3/2}}\cr&&\times\int\frac{d^3 p}
{(2\pi)^3}\bar\psi_{D(3/2)}\biggl({\bf p}+\frac{2\epsilon_q}{M_{D(3/2)}(w+1)}
{\bf \Delta}\biggr)\cr
&&\times\left[-2\epsilon_q
\overleftarrow{\frac{\partial}{\partial p}}+\frac{p}{\epsilon_q+m_q}\right]
\psi_B({\bf p}),\\
\label{tau1}
&&\!\!\!\!\!\!\!\tau_1(w)=\frac{\bar\Lambda'+\bar\Lambda}{w+1}\tau(w),\\
\label{tau2}
&&\!\!\!\!\!\!\!\tau_2(w)=-\frac{w}{w+1}(\bar\Lambda'+\bar\Lambda)\tau(w).
\end{eqnarray}

(ii) $B\to D^*_0e\nu$ and $B\to D^*_1e\nu$ decays
\begin{eqnarray}
\label{zeta}
&&\!\!\!\!\!\!\!\zeta(w)=\frac{\sqrt{2}}{3}\frac{1}{(w+1)^{1/2}}\cr
&&\times\int\frac{d^3 p}
{(2\pi)^3}\bar\psi_{D(1/2)}\left({\bf p}+\frac{2\epsilon_q}{M_{D(1/2)}(w+1)}
{\bf \Delta}\right)\cr
&&\times\left[-2\epsilon_q
\overleftarrow{\frac{\partial}{\partial p}}-\frac{2p}{\epsilon_q+m_q}\right]
\psi_B({\bf p}),\\
\label{zeta1}
&&\!\!\!\!\!\!\!\zeta_1(w)=\frac{\bar\Lambda^*+\bar\Lambda}{w+1}\zeta(w).
\end{eqnarray}
The contributions of all other subleading form factors, $\eta_i(w)$ and 
$\chi_i(w)$, to decay matrix elements are suppressed by an additional
power of the ratio $(w-1)/(w+1)$, which is equal to zero  at $w=1$ and
less than
$1/6$ at $w_{\rm max}=(1+r^2)/(2r)$.
Since the main contribution
to the decay rate comes from the values of form factors  close to $w=1$,
these form
factors turn out to be unimportant. This result is in  agreement with the
HQET-motivated 
considerations \cite{llsw} that the functions parametrizing  the
time-ordered products 
of the chromomagnetic term in the HQET Lagrangian with the leading
order currents should be small.     

The arrow over $\partial/\partial p$ in (\ref{tau}) and (\ref{zeta})  
indicates that the
derivative acts on the wave function of the $D^{**}$ meson. All the wave  
functions and
meson masses have been obtained in \cite{egf} by the numerical solution of the 
quasipotential equation. We use the following values for HQET parameters 
$\bar\Lambda=0.51$~GeV, $\bar\Lambda'=0.80$~GeV, and 
$\bar\Lambda^*= 0.89$~GeV \cite{egf}.

The last  terms in 
the square brackets of the expressions for the leading order Isgur--Wise  
functions 
$\tau(w)$ (\ref{tau}) and $\zeta(w)$ (\ref{zeta}) result from the wave\hfill 
function\hfill transformation\hfill (\ref{wig})\hfill 
associated \newline
with the relativistic  rotation of the
light quark spin (Wigner rotation) in
passing to the moving reference frame. These terms are numerically important
and lead to the suppression of the $\zeta$ form factor compared to 
$\tau$. Note that if we  
had applied a simplified non-relativistic quark model \cite{iw1,vo}
these important contributions would be missing. Neglecting further the
small difference between the wave functions $\psi_{D(1/2)}$ and 
$\psi_{D(3/2)}$, the following relation between $\tau$ and 
$\zeta$ would have been obtained \cite{llsw}
\begin{equation}\label{taunr}
\zeta(w)=\frac{w+1}{\sqrt{3}}\tau(w). 
\end{equation}
However, we see that this relation is violated if the relativistic
transformation properties of the wave function are taken into account. 
At the point $w=1$, where the initial $B$ meson and final $D^{**}$ are
at rest, we find instead the relation
\begin{equation}\label{diftau}
\frac{\tau(1)}{\sqrt{3}}-\frac{\zeta(1)}{2}\cong  
\frac12\int\frac{d^3p}{(2\pi)^3}
\bar\psi_{D^{**}}({\bf p})\frac{p}{\epsilon_q+m_q}\psi_B({\bf p}),
\end{equation}
obtained by assuming $\psi_{D(3/2)}\cong\psi_{D(1/2)}\cong\psi_{D^{**}}$.
The relation (\ref{diftau}) coincides with the one found in 
Ref.~\cite{mlopr}, where the Wigner rotation was also taken into account. 

In Table~\ref{tauv} we present our numerical results for the leading order
Isgur--Wise functions $\tau(1)$ and $\zeta(1)$  at zero recoil of the
final $D^{**}$ meson, 
as well as their slopes $\rho_{3/2}^2$ and $\rho_{1/2}^2$,
in comparison with other model predictions 
\cite{llsw,mlopr,ddgnp,w,cdp,gi,cccn}.  We see that most of 
the above approaches predict close values for the function $\tau(1)$ 
and its slope $\rho_{3/2}^2$, while the results for $\zeta(1)$ 
significantly differ from one another. This difference is a consequence
of a specific treatment of the relativistic quark dynamics. Nonrelativistic
approaches predict $\zeta(1)\simeq(2/\sqrt{3})\tau(1)$,
while the relativistic treatment leads to $(2/\sqrt{3}) \tau(1) > \zeta(1)$.
The more relativistic the light quark in the
heavy--light
meson is, the more suppressed  $\zeta$ is with respect to $\tau$.

\TABLE{
\caption{The comparison of our results for
the values of the leading Isgur--Wise functions $\tau$ and $\zeta$ at zero 
recoil of the final $D^{**}$ meson  and their slopes $\rho_j^2$ with other
predictions.} 
\label{tauv}
\begin{tabular}{cccccccc}
\hline
   & Ours & \cite{llsw} & \cite{ddgnp} & \cite{w} & \cite{cdp}& 
\cite{mlopr},\cite{gi} & \cite{mlopr},\cite{cccn}\\
\hline
$\tau(1)$ & 0.85 & 0.71 & 0.97 & 1.14 &     & 1.02 & 0.90\\
$\rho_{3/2}^2$  & 1.82 & 1.5  & 2.3  & 1.9  &     & 1.5  & 1.45\\
$\zeta(1)$ & 0.59 & 0.82 & 0.18 & 0.82 &$0.70\pm0.16$ & 0.44 & 0.12\\
$\rho_{1/2}^2$  & 1.37 & 1.0  & 1.1  & 1.4  &$2.5\pm1.0$ & 0.83 & 0.73\\ 
\hline
\end{tabular}
}

\TABLE{
\caption{Decay rates $\Gamma$ (in units  of $|V_{cb}/0.04|^2\times
10^{-15}$ GeV) 
and branching ratios BR (in \%) for  $B\to D^{**}e\nu$ decays in the
infinitely heavy quark
limit with the account of first  order $1/m_Q$ corrections. $R$ is
a ratio
of branching ratios with the account of  $1/m_Q$ corrections to branching
ratios in
the infinitely heavy quark limit.  }
\label{bd}
\begin{tabular}{cccccccc}
\hline
   &\multicolumn{2}{c}{$m_Q\to\infty$}&\multicolumn{2}{c}{With $1/m_Q$}&
&\multicolumn{2}{c}{Experiment}\\
Decay& $\Gamma$ & Br& $\Gamma$ & Br &$R$& Br (CLEO) \cite{cleo}  
& Br (ALEPH) \cite{aleph} \\
\hline
$B\to D_1e\nu$&1.4&0.32  & 2.7 & 0.63& 1.97
& $0.56\pm 0.13\pm0.08\pm0.04$& $0.74\pm0.16$\\
$B\to D^*_2e\nu$&2.1&0.51 & 2.5 & 0.59& 1.16& $<0.8$ & $<0.2$\\
$B\to D^*_1e\nu$&0.31&0.073 & 0.39 & 0.09&1.23 & &\\
$B\to D^*_0e\nu$&0.25&0.061 & 0.59 & 0.14&2.3 & &\\
\hline
\end{tabular}
} 

We can now calculate the decay  branching ratios by integrating double
differential  decay ra\-tes. Our results for decay rates both in
the infinitely heavy 
quark limit and taking account of the first order $1/m_Q$ corrections as well as
their ratio
$$R=\frac{{\rm Br}(B\to D^{**}e\nu)_{{\rm with}\, 1/m_Q}}{{\rm Br}(B\to
D^{**}e\nu)_{m_Q\to\infty}}$$
are presented in Table~\ref{bd}. We see that the inclusion
of $1/m_Q$
corrections considerably influences  the results and for some decays their
contribution
is as important as the leading order  contribution. This is the
consequence of the 
vanishing of the leading order contribution  to the decay matrix elements
due to the
heavy quark spin-flavour symmetry at zero recoil of the final $D^{**}$ meson
\cite{llsw}, while 
nothing prevents $1/m_Q$ corrections to  contribute to the decay matrix
element at 
this kinematical point. In fact, 
matrix elements at zero recoil are determined by the form factors
$f_{V_1}(1)$, 
$g_+(1)$ and $g_{V_1}(1)$, which receive non-vanishing contributions from
first order heavy quark mass corrections:
\begin{eqnarray}
\label{fv1}
\!\!\!\!\!\!\!\!\!\!\!\!\!\!\sqrt{6} f_{V_1}(1)&=&
-8\varepsilon_c(\bar\Lambda'-\bar\Lambda)\tau(1)\\
\label{g+}
\!\!\!\!\!\!\!\!\!\!\!\!\!\!g_+(1)&=&-\frac32(\varepsilon_c+\varepsilon_b)
(\bar\Lambda^*-\bar\Lambda)
\zeta(1)\\
\label{gv1}
\!\!\!\!\!\!\!\!\!\!\!\!\!\!g_{V_1}(1)&=&(\varepsilon_c-3\varepsilon_b)
(\bar\Lambda^*-\bar\Lambda)
\zeta(1).
\end{eqnarray}
Since the kinematically allowed range for  these decays is not broad
($1\le w\le 
w_{\rm max}\approx 1.32$), the contribution to the decay rate of the 
rather small $1/m_Q$ corrections is substantially increased \cite{llsw}. 
This is confirmed by
numerical calculations. From Table~\ref{bd} we see that
the decay
rate $B\to D^*_2e\nu$, for which all contributions  vanish at zero recoil,
is only slightly
increased by subleading $1/m_Q$ corrections. On the  other hand, $B\to
D_1e\nu$
and $B\to D^*_0e\nu$ decay rates receive large $1/m_Q$  contributions. The
situation
is different for the $B\to D^*_1e\nu$ decay. Here the  $1/m_Q$
contribution at zero recoil
is not equal to zero, but it is suppressed  by a very  small factor
$(\varepsilon_c-
3\varepsilon_b)$ (see Eq.~(\ref{gv1})), which is only  $\approx
0.015$~GeV$^{-1}$ for our
model parameters. As a result the $B\to D^*_1e\nu$ decay  rate receives
$1/m_Q$ 
contributions comparable  to those for the $B\to D^*_2e\nu$ rate. The
above discussion shows 
that the sharp increase  of $B\to D_1e\nu$ and $B\to D^*_0$ decay rates
by first order $1/m_Q$ corrections does not
signal the breakdown of  the heavy quark expansion, but is rather a result
of the
interplay of kinematical and dynamical effects. Thus we have  good reasons
to expect that higher  order $1/m_Q$ corrections will influence these
decay rates
at the level of 10 -- 20\%.  

In Table~\ref{bd} we present the experimental data from CLEO \cite{cleo}
and
ALEPH \cite{aleph}, which are  available only for the $B\to D_1 e\nu$
decay. For
$B\to D^*_2e\nu$, these experimental groups present only upper limits, which
require the use of some additional assumptions about the hadronic branching 
ratios of the
$D^*_2$ meson. Our result for  the branching ratio of the $B\to D_1e\nu$
decay with
the inclusion of $1/m_Q$ corrections  is in good agreement with both
measurements.
On the other hand, our branching ratio for the $B\to D^*_2e\nu$ decay is 
only within the CLEO upper
limit and disagrees with the ALEPH one.  However, there are some reasons
to expect 
that the ALEPH bound is too strong \cite{llsw}. 

Finally we test the fulfilment of the Bjorken 
sum rule \cite{b} in our model.
This sum rule states
\begin{equation}
\label{bsr}
\rho^2=\frac14+\sum_m\frac{|\zeta^{(m)}(1)|^2}{4}
 +2\sum_m \frac{|\tau^{(m)}(1)|^2}{3}
+\cdots ,
\end{equation}
where $\rho^2$ is the slope of the $B\to D^{(*)}e\nu$ Isgur--Wise function,
$\zeta^{(m)}$ and $\tau^{(m)}$  are the form factors describing the
orbitally excited states discussed above and their
radial excitations, and  
ellipses denote contributions from non-resonant channels. 
We see that the contribution of the lowest lying $P$-wave
states  implies the bound 
\begin{equation}
\rho^2>\frac14 +\frac{|\zeta(1)|^2}{4}+2\frac{|\tau(1)|^2}{3}=0.81,
\end{equation}  
which is in agreement with the slope $\rho^2=1.02$ in our model \cite{fg}
and with experimental values \cite{pdg}.

\section{Semileptonic decays to radially excited states}
 
In the case of semileptonic $B$ decays to radially excited $D$ mesons we 
get the following
expressions for leading and subleading Isgur-Wise functions \cite{rad}:
\begin{eqnarray}
\label{xi}  
&&\!\!\!\!\!\xi^{(1)}(w)=\left(\frac{2}{w+1}\right)^{1/2}\cr
&&\!\!\!\!\!\times\!\int\!\!\frac{d^3 p}
{(2\pi)^3}\bar\psi^{(0)}_{D^{(*)}{'}}\!\!\left({\bf p}+
\frac{2\epsilon_q}{M_{D^{(*)}{'}}(w+1)}
{\bf \Delta}\right)
\psi^{(0)}_B({\bf p}),\cr&&\\ 
\label{xi3}
&&\!\!\!\!\!\tilde\xi_3(w)=\left(\frac{\bar\Lambda^{(1)}+\bar\Lambda}2-m_q+
\frac16\frac{\bar\Lambda^{(1)}-\bar\Lambda}{w-1}\right)\cr
&&\!\!\!\!\!\times\left(1+
\frac23\frac{w-1}{w+1}\right)\xi^{(1)}(w),\\ 
\label{chi1}
&&\!\!\!\!\!\tilde\chi_1(w)\cong \frac1{20}\frac{w-1}{w+1}\frac{\bar\Lambda^{(1)}
-\bar\Lambda}{w-1}\xi^{(1)}(w)\cr 
&&\!\!\!\!\!+\frac{\bar\Lambda^{(1)}}2
\left(\frac{2}{w+1}\right)^{1/2}\cr
&&\!\!\!\!\!\times\!\int\!\!\frac{d^3 p}
{(2\pi)^3}\bar\psi^{(1)si}_{D^{(*)}{'}}\!\!\left({\bf p}+
\frac{2\epsilon_q}{M_{D^{(*)}{'}}(w+1)}
{\bf \Delta}\right)
\psi^{(0)}_B({\bf p}),\cr&&\\ 
\label{chi2}
&&\!\!\!\!\!\tilde\chi_2(w)\cong -\frac1{12}\frac{1}{w+1}\frac{\bar\Lambda^{(1)}
-\bar\Lambda}{w-1}\xi^{(1)}(w),\\ 
\label{chi3}
&&\!\!\!\!\!\tilde\chi_3(w)\cong -\frac3{80}\frac{w-1}{w+1}\frac{\bar\Lambda^{(1)}
-\bar\Lambda}{w-1}\xi^{(1)}(w)\cr 
&&\!\!\!\!\!+\frac{\bar\Lambda^{(1)}}4
\left(\frac{2}{w+1}\right)^{1/2}\cr
&&\!\!\!\!\!\times\!\int\!\!\frac{d^3 p}
{(2\pi)^3}\bar\psi^{(1)sd}_{D^{(*)}{'}}\!\!\left({\bf p}+
\frac{2\epsilon_q}{M_{D^{(*)}{'}}(w+1)}
{\bf \Delta}\right)
\psi^{(0)}_B({\bf p}),\cr&&\\ 
\label{chib}
&&\!\!\!\!\!\chi_b(w)\cong \bar\Lambda\left(\frac{2}{w+1}\right)^{1/2}\cr
&&\!\!\!\!\!\times\!\int\!\!\frac{d^3 p}
{(2\pi)^3}\bar\psi^{(0)}_{D^{(*)}{'}}\!\!\biggl({\bf p}+
\frac{2\epsilon_q}{M_{D^{(*)}{'}}(w+1)}
{\bf \Delta}\biggr)\cr
&&\!\!\!\!\!\times\left[\psi^{(1)si}_B({\bf p})-3\psi^{(1)sd}_B({\bf p})\right],
\end{eqnarray}
where ${\bf \Delta}^2=M_{D^{(*)}{'}}^2(w^2-1)$.
Here we used the expansion for the $S$-wave meson wave function
$$\psi_M=\psi_M^{(0)}+\bar\Lambda_M\varepsilon_Q\left(\psi_M^{(1)si}
+d_M\psi_M^{(1)sd}\right)+\cdots,$$
where $\psi_M^{(0)}$ is the wave function in the limit $m_Q\to\infty$,
$\psi_M^{(1)si}$ and $\psi_M^{(1)sd}$ are the spin-independent and 
spin-dependent first order $1/m_Q$ corrections, $d_P=-3$ for pseudoscalar and
$d_V=1$ for vector mesons.
The symbol $\cong$ in the expressions (\ref{chi1})--(\ref{chib}) for the
subleading functions $\tilde\chi_i(w)$ means that the corrections
suppressed by an additional power of the ratio $(w-1)/(w+1)$, which is equal 
to zero  at $w=1$ and less than $1/6$ at $w_{\rm max}$, were neglected. 
Since the main contribution to the decay rate comes from the values of 
form factors  close to $w=1$, these corrections turn out to be unimportant. 
 
It is clear from the expression (\ref{xi}) that the leading order contribution
vanishes at the point of zero recoil (${\bf \Delta}=0, w=1$) of the 
final $D^{(*)}{'}$ meson, 
since the radial parts of the wave functions $\Psi_{D^{(*)}{'}}$ and $\Psi_B$
are orthogonal in the infinitely heavy quark limit. The $1/m_Q$ 
corrections to the current (\ref{corc}) also do not contribute at this 
kinematical point for the same reason.
The only nonzero contributions at $w=1$  come from corrections to the
Lagrangian~\footnote{There are no normalization conditions for 
these corrections
contrary to the decay to the ground state $D^{(*)}$ mesons, where the 
conservation of vector current requires their vanishing at
zero recoil \cite{luke}.} $\tilde\chi_1(w)$, $\tilde\chi_3(w)$ and 
$\chi_b(w)$. From Eqs.~(\ref{cff}) one can find for the form factors
contributing to the decay matrix elements at zero recoil 
\begin{eqnarray}\label{zeroc}
\!\!\!\!\!\!\!\!\!\!h_{+}(1)&=&\varepsilon_c\left[2\tilde\chi_1(1)+
12\tilde\chi_3(1)\right]+\varepsilon_b\chi_b(1),\cr
\!\!\!\!\!\!\!\!\!\!h_{A_1}(1)&=&\varepsilon_c\left[2\tilde\chi_1(1)-4\tilde\chi_3(1)\right]
+\varepsilon_b\chi_b(1).
\end{eqnarray}
Such nonvanishing contributions at zero recoil   
result from the first order $1 /m_Q$ corrections to the wave functions (see 
Eq.~(\ref{chib}) and the last terms in
Eqs.~(\ref{chi1}), (\ref{chi3})). Since the kinematically
allowed range for these decays is not broad ( $1\le w\le 
w_{\rm max}\approx 1.27$) the relative contribution to the decay rate of such 
small $1/m_Q$ corrections is substantially increased.  
Note that the terms 
$\varepsilon_Q(\bar\Lambda^{(n)}-\bar\Lambda)\xi^{(n)}(w)/(w-1)$ have
the same behaviour near $w=1$ as the leading order contribution, in contrast
to decays to the ground state $D^{(*)}$ mesons, 
where $1/m_Q$ corrections 
are suppressed with respect to the leading order contribution 
by the factor $(w-1)$ near this point (this result is known as Luke's 
theorem \cite{luke}). Since inclusion of first order heavy quark corrections to
$B$ decays to the ground state $D^{(*)}$ mesons results in approximately a 
10-20\% increase of decay rates \cite{fg,n}, one could expect that the 
influence of these corrections on decay rates to radially excited 
$D^{(*)}{'}$ mesons will be more essential. Our numerical analysis supports
these observations.

\TABLE{
\caption{Decay rates $\Gamma$ (in units  of $|V_{cb}/0.04|^2\times
10^{-15}$ GeV) 
and branching ratios BR (in \%) for  $B$ decays to radially
excited $D^{(*)}{'}$  mesons in the
infinitely heavy quark
limit and taking account of first  order $1/m_Q$ corrections. 
$\Sigma (B\to D^{(*)}{'}e\nu)$ 
represent the sum over the channels. 
$R'$ is a ratio
of branching ratios taking account of  $1/m_Q$ corrections to branching
ratios in
the infinitely heavy quark limit.  }
\label{tbr}
\begin{tabular}{cccccc}
\hline
   &\multicolumn{2}{c}{$m_Q\to\infty$}&\multicolumn{2}{c}{With $1/m_Q$}\\
Decay& $\Gamma$ & Br& $\Gamma$ & Br &$R'$ \\
\hline
$B\to D'e\nu$&0.53&0.12  & 0.92 & 0.22& 1.74\\
$B\to D^{*}{'}e\nu$&0.70&0.17 & 0.78 & 0.18& 1.11\\
$\Sigma (B\to D^{(*)}{'}e\nu)$ & 1.23& 0.29&1.70&0.40&1.37\\
\hline
\end{tabular}
} 

We can now calculate the decay  branching ratios\hfill by\hfill integrating\hfill double\hfill
differential\hfill  decay\newline rates. Our results for decay rates
both in the infinitely heavy quark limit and taking account of the 
first order $1/m_Q$ corrections as well as their ratio
$$R'=\frac{{\rm Br}(B\to D^{(*)}{'}e\nu)_{{\rm with}\, 1/m_Q}}{{\rm Br}(B\to
D^{(*)}{'}e\nu)_{m_Q\to\infty}}$$
are presented in Table~\ref{tbr}. 
We find that both $1/m_Q$ corrections to decay
rates arising from corrections to HQET Lagrangian (\ref{chi1})--(\ref{chib}),
which do not vanish at zero recoil, and corrections to the current (\ref{xi3}),
(\ref{corc}), vanishing at zero recoil, give significant contributions. In
the case of $B\to D'e\nu$ decay both types of these corrections tend to
increase the decay rate leading to approximately a 75\% increase of the
$B\to D'e\nu$ decay rate. On the other hand, these corrections give opposite
contributions to the $B \to D^*{'}e\nu$ decay rate: the corrections to the
current give a negative contribution, while  corrections to the Lagrangian
give a positive one of approximately the same value. This interplay of
$1/m_Q$ corrections only slightly ($\approx 10\%$) increases the decay rate
with respect to the infinitely heavy quark limit. As a result the branching
ratio for $B\to D'e\nu$ decay exceeds the one for $B\to D^*{'}e\nu$
after inclusion of first order $1/m_Q$ corrections. In the infinitely heavy
quark mass limit we have for the ratio $Br(B\to D'e\nu)/Br(B\to D^*{'}e\nu)
=0.75$, while the account of $1/m_Q$ corrections results in the considerable 
increase of this ratio  to 1.22.

In Table~\ref{tbr} we also present the sum of the branching ratios over first
radially excited states. Inclusion of $1/m_Q$ corrections results in
approximately a 40\% increase of this sum. We see that our model predicts
that $ 0.40\%$ of $B$ meson decays go to the first radially excited $D$
meson states. If we add this value to our prediction for $B$ decays to
the first orbitally excited states $ 1.45\%$ \cite{orb}, we 
get the value of 1.85\%. This result means that approximately 2\% of
$B$ decays should go to higher charmed excitations.   

\section{Conclusions}
In  this paper we have applied the relativistic quark model to the
consideration
of  semileptonic $B$ decays to orbitally and radially excited 
charmed mesons, in the leading
and  subleading orders of the heavy quark expansion.  We have found an
interesting 
interplay  of relativistic and finite heavy quark mass contributions.
In particular,
it has been  found that  the Lorentz transformation properties of meson
wave functions 
play an  important role in the theoretical description of these decays. 
Thus, the  Wigner rotation of the light quark spin gives a significant
contribution 
already at  the leading order of the heavy quark expansion for decays
to orbitally excited mesons. This contribution
considerably  reduces the leading order Isgur--Wise function $\zeta$ with
respect to
$\tau$. As a  result, in this limit the decay rates of $B\to D^*_0e\nu$ and
$B\to D^*_1e\nu$     
are approximately  an order of magnitude smaller than the decay rates of
$B\to D_1e\nu$ and 
$B\to D^*_2e\nu$. On  the other hand, inclusion of the first order $1/m_Q$
corrections
also substantially  influences the decay rates. This large effect of
subleading heavy
quark corrections is  a consequence of vanishing the leading order
contributions
to the decay matrix  elements due to heavy quark spin-flavour symmetry at
the point
of zero recoil of  the final charmed meson. However, the subleading order
contributions do not vanish at this
kinematical point.  Since the kinematical range for these decays is rather
small, the
role of these  corrections is considerably increased. Their account
results in
an approximately  twofold enhancement of the $B\to D_1e\nu$, $B\to
D^*_0e\nu$ and $B\to D'e\nu$
decay rates, while  the  $B\to D^*_2e\nu$, $B\to D^*_1e\nu$ and $B\to D^{*}{'}e\nu$
rates are increased only slightly. The small influence of $1/m_Q$ corrections on the
latter decay rate is the consequence of the additional interplay of
$1/m_Q$  corrections. We thus see
that these subleading heavy quark corrections turn out to be very important
and considerably change results in the infinitely heavy quark limit. 
For example, the ratio
of branching ratios  ${\rm Br}(B\to D^*_2e\nu)/{\rm Br}(B\to D_1e\nu)$
changes from
the value of about 1.6 in the heavy quark limit, $m_Q\to\infty$,
to the value of about 1 after subleading corrections are included.
Finally we find that the semileptonic $B$ decays to first orbital and 
radial excitations of $D$ mesons amount in total to approximately
2\% of the $B$ decay rate.      
 
We are grateful to the organizers for the nice meeting and stimulating discussions.
Two of us (R.N.F and V.O.G.) were supported in part  by {\it Russian Foundation for
Fundamental Research} under Grant No.\ 00-02-17768.

\end{document}